# Immigrant and native's export benefiting from business collaborations: a global study

## Shayegheh Ashourizadeh*

School of Economics and Management,
Tsinghua University,
Beijing, China
Email: shaya@sem.tsinghua.edu.cn
*Corresponding author

## Mehrzad Saeedikiya

Faculty of Political, Economic and Social Sciences,
University of Milan,
Milan, Italy
and
School of Economics and Management,
Tsinghua University,
Beijing, China
Email: Mehrzad.saeedikiya@studenti.unimi.it

**Abstract:** The authors hypothesised that export develops in the network of business collaborations that are embedded in migration status. In that, collaborative networking positively affects export performance and immigrant entrepreneurs enjoy higher collaborative networking than native entrepreneurs due to their advantage of being embedded in the home and the host country. Moreover, the advantage of being an immigrant promotes the benefits of collaborative networking for export compared to those of native entrepreneurs. A total of 47,200 entrepreneurs starting, running and owning firms in 71 countries were surveyed by Global Entrepreneurship Monitor and analysed through hierarchical linear modelling technique. Collaborative networking facilitated export and migration status influenced entrepreneurs' networking, in that, immigrant entrepreneurs had a higher level of collaborative networking than native entrepreneurs. Consequently, immigrant entrepreneurs seemed to have benefited from their network collaborations more than their native counterparts did. This study sheds light on how immigrant entrepreneurs' network collaborations can be effective for their exporting.

**Biographical notes:** Shayegheh Ashourizadeh received her PhD in Business Economics in 2017 from the University of Southern Denmark. She has been affiliated with the Sino-Danish Center for Education and Research from 2014 to 2017. Currently, she is a Postdoctoral Fellow at Tsinghua University, China.

She researches the endeavours of nascent entrepreneurs as embedded in institutions, utilising data from the Global Entrepreneurship Monitor. She has presented her works at international conferences, including RENT (2012–2018), Babson (2016) and Druid (2016). Her recent works are published in international journals including *European Journal of International Management*, *International Journal of Business and Globalisation*, *International Journal of Entrepreneurship and Small Business*, and *Gender in Management*.

Mehrzad Saeedikiya is an early career researcher of entrepreneurship and labour studies at University of Milan, Italy. He is a visiting researcher at Department of Strategy, Entrepreneurship and Innovation at Tsinghua University, China. He is interested in researching entrepreneurship and innovation in the context of institutions.

# 1 Introduction

To study the internationalisation of entrepreneurial firms, it is suggested that the analysis be expanded beyond the firm strategies to address the impact of a firm's role and position within a network of relationships (Hoang and Antoncic, 2003 Coviello, 2006). This need arises because the internationalisation of the firm is not only the result of the strategic decisions of the firm's managers, but also the result of opportunities created by the firm's network (Coviello and Munro, 1995). This method is called the network approach based on its use several studies have focused on the role of networks in the internationalisation of the firms (e.g. Ciravegna et al., 2014a; Ciravegna et al., 2014b; Cavusgil and Knight, 2015; Hohenthal et al., 2015; Oparaocha, 2015). The basic argument of such studies is that networks provide entrepreneurs with valuable knowledge, capital, advice and support for their exporting behaviour (Portes, 1995; He and Wei, 2013; Elo, 2015; Oparaocha, 2015; Wąsowska et al., 2016).

Accordingly, one area of these studies has been devoted to investigating the factors changing the content and the structure of the entrepreneurs' networks and exploring how exporting develops in those networks (e.g. Jensen and Schøtt, 2015; Schøtt, 2017, 2018). Migration status is an important factor which reflects the concept of being embedded in two societies and, based on this, entrepreneurs can be categorised as either immigrants or natives (White and Tadesse, 2008; Mustafa and Chen, 2010; Chung and Tung, 2013). Entrepreneurs are considered as natives if they pursue their entrepreneurial activity in their country of birth, while if they engage in this in a country other than that of their birth, they are recognised as immigrant entrepreneurs (Global Entrepreneurship Monitor, 2013).

Immigrant entrepreneurs represent a unique case for study to gain knowledge about the impact of networking on exporting because they are not only embedded in two societies, namely, home and host country, but also in two worlds of connections in either private social relations with friends and families or public business relations with co-workers and other entrepreneurs (Schøtt, 2016, 2017). This dual embeddedness, that is, their integration in two societies (home and host) and two worlds of connections requires them to simultaneous play the two interrelated roles of immigrant and entrepreneur. Through their immigrant role, immigrant entrepreneurs are embedded in their ethnic and private social ties in the host society while keeping, cutting, weakening or re-considering their former social relationships in their home country (Drori et al., 2009). While they have the opportunity for getting free and reliable information, consultancy and support from their social network for exporting (Majocchi et al., 2016), their embeddedness in two societies may also put them in a superior position to that of the natives for accessing international markets (Light and Shahlapour, 2016). They can act as the connector of the two markets (Kloosterman, 2010; Aleksynska and Peri, 2014, Emontspool and Servais, 2017; Griffin-El and Olabisi, 2018) and as the receiver of more heterogeneous ideas for exporting via their being embedded in different worlds of connections (Aleksynska and Peri, 2014). Such a theoretical stance has been supported by empirical evidence which shows that the immigrant entrepreneurs can benefit from these social networks more than the natives when it comes to exporting (Portes, 1995; Zaheer et al., 2009; Jean et al., 2011; Ashourizadeh et al., 2016; Wales et al., 2016).

Immigrant entrepreneurs are not only embedded in their private and ethnic social relations as a result of their migration, but as entrepreneurs, they are also embedded in their business relations. Generally, social relationships bring value, such as knowledge and information, financial or emotional support, in form of social capital for individuals (Payne et al., 2011). The content and the structure of their business collaborations and its benefit for exporting can be affected by the limitations imposed or opportunities created through the migration process. On one hand, they have some immediate disadvantages in comparison with the natives and this makes it essential for them to widen or intensify their collaboration networks to balance these drawbacks. For example, they may bring some competencies which are not directly transferable to the requisites of the host society (Aliaga-Isla and Rialp, 2013; Chiesi, 2014) and so need to balance this limitation with networking with different economic contracts. They also need local knowledge about the host market and the conditions for doing business along with the procedures and regulations which can lead to them having more collaborative networking with their business partners than the natives do. On the other hand, as an advantage, immigrant entrepreneurs may have pre-existing ties in global networks, know foreign languages and have cross-cultural competences (Portes et al., 2002; Coe and Bunnell, 2003; Rusinovic, 2008; Kloosterman, 2010; Cerdin et al., 2014). These qualities are not only effective in extending business collaborations into the international arena, but they may also lead to the development of export in such business collaboration networks.

Notwithstanding the foregoing arguments present in the literature, there is a paucity of research on the content and structure of the business collaboration networks of immigrant and native entrepreneurs and their effect on export. This current study is an attempt to address this paucity. Therefore, it set out to investigate the business relationships of immigrant entrepreneurs and the benefit of such connections for exporting when compared to those for the natives. To accomplish this, the following research questions were asked:

- *How different is the business collaboration network among immigrant entrepreneurs as compared to the natives?*

- *How different is the benefit of business collaboration network for exporting among immigrant entrepreneurs as compared to the natives?*

The contribution of this study is that it expands the application of the concept of embeddedness beyond that of the economic activity in kin or ethnic social relations. Instead, it supports the idea of mixed embeddedness in the sense that, economic activity is not only embedded in the close kin or ethnic ties of the individuals, but also in the social ties in the economic and institutional context of the host society (Kloosterman et al., 1999). By investigating a specific type of immigrant network that goes beyond strong ties of family and friends, this study sought to identify evidence on the embeddedness of the economic activity (here, exporting) in the business collaborations network of the immigrant and native entrepreneurs. Further, the findings also contribute to the existing theory on how the dual embeddedness of immigrants' networking in their home and host countries affect their benefit for exporting. In addition to this, due to the scope of the population sampled, the findings can be generalised to the population of immigrant and native entrepreneurs worldwide.

This paper is organised as follows. First, there is the theoretical framework for making hypothetical predictions about the effect of migration status on collaborative networking and its benefit for exporting. In the next section, the research design and methodology are presented, and later on, the results are presented. Finally, there are the summary and discussion of the findings from the research, its contribution to theory and the implications for future research.

## 2 Theoretical background and hypotheses

In this section, we review the effect of migration status upon collaborative business networking followed by an examination of the effects of migration status and collaborative networking on export. Finally, we study the moderating effect of migration status on the benefits of business collaboration network for export.

### 2.1 Business collaboration network affected by migration status

An immigrant is 'a person who lives temporarily or permanently in a country where he or she was not born and has acquired some significant social ties to this country' (UNESCO, 2016). Accordingly, an immigrant entrepreneur is defined as a person who lives in a country other than the one in which s/he was born and pursues an entrepreneurial activity in that country (Global Entrepreneurship Monitor, 2013). In light of these definitions, the migration status of an entrepreneur is the condition of his or her being either immigrant or native. Immigrant entrepreneurs are dually embedded in two societies and two worlds of connections (Drori et al., 2009; Schøtt, 2017) that can bring some advantages over the natives in terms of doing business in the host society. It has been argued that their embeddedness in two societies have some advantages for them in terms of getting advice as well as access to cheap labour and sources of finance for entrepreneurial activity (Basu and Altinay, 2002; Peters, 2002; Portes and Shafer, 2007; Ashourizadeh et al., 2016). In a study by Schøtt (2018), it was shown that Caribbean entrepreneurs in the diaspora had more business collaboration than those who resided in their home countries. The author also provided empirical evidence in support of the benefit of such transnational network for innovation, growth and exporting. These refer to the collaborations of a firm with the actors of the market, including suppliers, customers, distributors, competitors and government (Johanson and Vahlne, 2009; Johanson and Mattson, 2015).

Other researchers investigated the greater possibility of immigrant entrepreneurs establishing international business relations than the natives (Drori et al., 2009; Patel and Conklin, 2009). When compared to the natives who often lack language skills, cultural know-hows and contacts for building business relationships in the destination market (Portes et al., 2002; Saxenian, 2002; Cerdin et al., 2014), immigrant entrepreneurs seemed to have the greater ability to establish business collaborations with their home country actors. The Taiwanese and Indian entrepreneurs of Silicon Valley who have connected Asia Pacific technology companies and technical expertise to the USA are good examples of this comparative advantage (Saxenian, 2002). In both cases, collaborative business networking with their home market actors has helped them to share and acquire manufacturing capabilities and skills.

Additionally, collaborative business networking as a competitive strategy seems more crucial for the immigrants than the natives. Immigrant entrepreneurs can rely on their business collaboration for gaining market knowledge, mitigating the risk, strengthening their brand, and increasing the scale and speed of accessing the market (Hernandez-Espallardo et al., 2018). Compared to the natives, immigrant entrepreneurs encounter more asymmetric information about the host country's market. Typically, the imbalance of the knowledge between the two parties creates asymmetric information and a competitive advantage for the party who has better or more knowledge (Akerlof, 1978). Compared to the immigrant entrepreneurs, natives have a greater knowledge of their home country and its market which allows them to share or sell this kind of information to the immigrants in different forms of business contracts, such as direct selling and business partnerships or collaborations.

Since business networking can support immigrant entrepreneurs in case of general supply fluctuations, labour supply changes and changes in substitute products, markets and the values of consumers (Hernandez-Espallardo et al., 2018), immigrant entrepreneurs may benefit from these networks for mitigating the risk. Additionally, as a strategy for strengthening their brand and for better penetration of the market, immigrant entrepreneurs need to partner with trustworthy firms in the host market. The expansion of the business networks with reliable market actors can create credit and reputation for them and signal their value and reliability to the market (Hoang and Antocic, 2003). Based on the foregoing arguments the authors hypothesise that:

*H1: Immigrant entrepreneurs have more collaborative business networking than native entrepreneurs.*

### 2.2 Export affected by business network collaborations

Network theory provides useful insights into the dynamics of a firm's internationalisation by examining a variety of issues, including the impact of network relationships on foreign market selection and the relative influence of other firms on new market entry strategies (Coviello and Munro, 1995). Based on this perspective, entry to international markets is not solely the result of the managers' strategic decision, but also of the opportunities created through entrepreneurial networks (Coviello and Munro, 1995; Hoang and Antoncic, 2003).

In the context of entrepreneurship, considerable attention has been given to the effect of ethnic and social networks on the entry into entrepreneurial activity and post-entry performance. Such networks provide information on market opportunities for starting a new business (Chung and Dahms, 2016) or consultancy and emotional support for risk-taking in business startups or failures (Gimeno et al., 1997; Brüderl and Preisendörfer, 1998). For exporting, ethnic and social networks provide information on export destinations and opportunities (Ashourizadeh and Schøtt, 2016; Emontspool and Servais, 2017). Similarly, they act as a source of personal advice, reliable consultancy and cheap labour (Chand and Ghorbani, 2011).

Similar to the kin or ethnic social networks, a firm's collaborative networking with business partners can also play an important role in its internationalisation in three different ways. First, it can provide the firm with the information and advice necessary for remaining competitive in the market. Venture capitalists and professional service organisations are a part of these business collaborations providing information on the market (Hoang and Antoncic, 2003; Forsgren and Johanson, 2014). Second, business collaborations can provide legitimacy and credit to a firm by a reputational and signalling mechanism that assists in reducing the risk perceived by resource holders. This is accomplished by their association with well-regarded individuals and organisations (Hoang and Antoncic, 2003; Ahlers et al., 2015). Third, the networks can increase the amount of resources available for doing entrepreneurial activity and the speed of accessing them (Hoang and Antoncic, 2003; Forsgren and Johanson, 2014; Knight and Liesch, 2016). There are several empirical studies which advocate this theoretical perspective. Jensen and Schøtt (2014) examined the effect of network collaboration on entrepreneurs' performance in the development of innovation in China by comparing the country as a fast-growing economy with the rest of the world. The results of the research indicated that relationships in the business sphere enhanced the entrepreneurs' innovations. The results of the study were confirmed in another one which examined the influence of the firm's network on innovation globally (Schøtt and Jensen, 2016). Furthermore, an empirical study by Ashourizadeh and Schøtt (2016) also established that entrepreneurs' performance in export was influenced by their network, mainly in the public sphere. Repeatedly, it was shown that entrepreneurs' networking in the public sphere enhanced their export (ibid).

To summarise these findings, collaborative business networking can generally influence the internationalisation of the firm by playing a facilitating and signalling function. Through the facilitating function, the networks provide or accelerate access to intangible resources like information and market advice, and tangible ones such as raw materials and labour. Both of these kinds of resources are useful for exporting. Similarly, through their signalling function, business collaborations reflect the credibility and reliability of the firms in their circle to the other market actors, including venture capitalists and well-established distribution channels in the destination market for accessing international markets. Based on these arguments, the authors hypothesise that:

*H2: Business collaborations positively affect exporting, in the way that the more collaborative business networking the more export.*

### 2.3 The moderating effect of migration status on the benefit of business collaboration on export

Immigrant entrepreneurs are embedded in two societies and in their worlds of connections. As they are the connector in at least two societies, they are situated in such a way that they receive information from different sources which can help them to develop export ideas and to know about export opportunities. Information is more homogeneous among individuals belonging to a single group and as the number of the groups rises, the heterogeneity of the information increases (*Burt, 2002, 2009*). Immigrant entrepreneurs are not only embedded in their ethnic and kin ties, but also in their host society's contacts and collaborations. They gain certain advantages from the way they are positioned or embedded in these social structures (Borgatti, 1997; Burt, 2009). Due to their location the transitional points (Burt, 2009) and being embedded in two different societies (Drori et al., 2009), immigrant entrepreneurs may acquire more heterogeneous and complementary information from their former and new ties. For exporting, immigrants have the possibility of combining ideas coming from different sources and use them either in developing export ideas or in recognising export destinations, conditions and possibilities.

Compared to the natives, immigrants have cross-cultural competencies, know the language and possess the know-hows of their home society. All of these can be beneficial for them for accessing the international market and evaluating its demands for export (Coe and Bunnell, 2003; Rusinovic, 2008; Kloosterman, 2010). Another source of export benefit for the immigrant entrepreneurs is the support and business strategies established by their ethnic community. For example, local ethnic professional associations can provide the recently arrived immigrant entrepreneurs with the resources and contacts (Saxenian, 2002). This would be in addition to their business activity in the host society which can lead to business collaborations. These social arrangements act as enablers which help even the new immigrant entrepreneurs to locate and maintain collaborations across long distances and gain access the resources, production facilities, skills and markets (Coe and Bunnell, 2003). For example, immigrant entrepreneurs in the field of technology can benefit from subcontracting their manufacturing to China and Taiwan and their software design and development to India (Saxenian, 2002). As a result, while they are in the process of establishing their business collaborations, they are doing the same for the internationalisation of their business. Based on the foregoing arguments, the authors hypothesise that:

*H3: Immigrant entrepreneurs benefit from their business collaboration network for exporting more than native entrepreneurs.*

## 3 Research design and data

The hypotheses concern the effect of migration status and business relations upon export in a population of entrepreneurs owning, managing, starting or operating companies in 71 countries around the world. Therefore, the dependent variable is export and the independent variables are migration status and business relations.

### 3.1 The sample

The data used in this paper were collected by Global Entrepreneurship Monitor (GEM) which conducts an annual survey by having participating countries assess adult involvement in entrepreneurship (Reynolds et al., 2005). GEM utilises two-stage sampling, the first of which is the sampling of the countries followed by that of the adults. The countries are self-selected and usually form a national team that joins GEM in order to record the entrepreneurs' aspirations and performance, including their export. This was done in 71 countries including Algeria, Angola, Argentina, Austria, Barbados, Belgium, Bosnia and Herzegovina, Botswana, Brazil, Chile, China, Colombia, Costa Rica, Croatia, Denmark, Ecuador, Egypt, El Salvador, Estonia, Ethiopia, Finland, France, Germany, Ghana, Greece, Hungary, India, Indonesia, Iran, Ireland, Israel, Italy, Jamaica,

Japan, Korea, Latvia, Lithuania, Macedonia, Malawi, Malaysia, Mexico, Namibia, Netherlands, Nigeria, Norway, Pakistan, Palestine, Panama, Peru, Philippines, Poland, Portugal, Puerto Rico, Romania, Russia, Singapore, Slovakia, Slovenia, South Africa, Spain, Sweden, Switzerland, Taiwan, Thailand, Trinidad and Tobago, Tunisia, Turkey, Uganda, the UK, Uruguay, Vietnam and Zambia. In 2012, the Adult Population Survey (APS) included questions for owner-managers about their business relations and also their migration status in each country (Global Entrepreneurship Monitor, 2013). The sample for this study was based on data from this source and consisted of 47,200 entrepreneurs who were starting, running or owning a firm who were asked about their migration status, business relations and export activities.

GEM data can be useful for research for several reasons. First, GEM uses a standardised approach in its attempt to collect fairly comprehensive information about entrepreneurship in different countries and types of economies. In addition to this, the definition for entrepreneurship used in the GEM survey is in line with that of key scholars on entrepreneurship which is viewed as a process (Shane and Venkataraman, 2000). As a result, the GEM survey tends to cover some of the key elements of the entrepreneurship process, such as the motive to start a business and the individual company's characteristics and performance (Bosma et al., 2012). Furthermore, the GEM Adult Population Survey provides the statistical characteristic at the country level. Finally, the annual survey (from 2001 to 2016) conducted in more than one hundred countries, with at least 2000 samples per economy, results in a massive dataset for research in entrepreneurship and adds to the generalisability of the results (Bosma, 2013).

## 3.2 Measurements

### 3.2.1 Entrepreneurs' export

There are national data on exports held by organisations such as the World Trade Organization (WTO), the United Nations (UN) and the Organization of Economic Co-operation and Development (OECD), but they do not include export reports from individual firms. However, GEM collects data on the exporting activities of individual firms and so this source was used in this study and is consistent with previous studies on export using this type of measurement (De Clercq et al., 2008; Hessels and van Stel, 2011). In this study, a firm's export activity is normally assessed based on the proportion of customers outside the country and is obtained by asking: 'What proportion of your customers normally lives outside your country? – Is it more than 90%, more than 75%, more than 50%, more than 25%, more than 10% or 10% or less?' The answers are then coded as a percentage of the median of the interval for customers outside the country as: 0%, 5%, 17%, 37%, 62%, 82% and 95% and then logged to reduce the skewness of numbers and gain a normal distribution among the cases.

Since our analysis consists of starting and operating firms, responding about the number of customers is more convenient for owner managers of starting firms (Liu and Schøtt, 2019). Though, asking about export turnover share is more typical in surveys.

### 3.2.2 Migration status

In the United Nations Educational, Scientific and Cultural Organization (UNESCO), the term 'immigrant' is defined as 'any person who lives temporarily or permanently in a country where he or she was not born and has acquired some significant social ties to this country' (see UNESCO, 2016). In a similar vein, GEM distinguishes immigrants from natives via their status, asking this question 'Were you born in this country?' If the respondent replies 'Yes, I was born in this country', then the answer is coded as 0 and this means s/he is regarded as a native. If the respondent replies 'No, I was not born in this country', then the answer will be coded as 1 which means s/he is regarded as an immigrant.

The variable of migration status, i.e. native versus immigrant entrepreneurs, is only a sketch of a more complicated reality. Here, this variable ignores the status of being adopted or born in a parent home country and then being brought to the host country from childhood, so there is a lack of clarity regarding this aspect. However, GEM has the advantage of collecting information about immigrants and natives at the same point in time on a large scale from various ethnicities in the host countries at different stages of economic development stage. This has hardly been done in other surveys and the data have been accepted worldwide for use by the academic community (Liu and Schøtt, 2019).

### 3.2.3 Collaborative business networking

The entrepreneurs are embedded in their business relations which are distinguished by the nature of their collaborations. GEM has measured such types of collaborations by asking: *Does your business collaborate with other enterprises or organisations to produce goods or services? Does your business collaborate with other enterprises or organisations to procure supplies? Is your business with others to sell your products or services to your current customers? Is your business working together with others to sell your products or services to new customers? Is your business working together with others to create new products or services to your current customers? Is your business working together with others to create new products or services to new customers? Is your business working together with others about how to make your business more effective?* Each question requires a 'Yes' or 'No' response with 'Yes' being coded as 1 and 'No' as 0.

If the answer is 'Yes', a follow-up question is then asked to find out the extent of the collaboration in terms of whether or not it was a committed and close relation, the frequency of contact and the amount of investment of effort, energy and time in such a business relationship (Jensen and Schøtt, 2014). Therefore, through these questions, not only did the researchers obtain information regarding the absence or presence of a business relationship, but also about the degree of intensity measured on a scale of 0 for no business relationship, 0.5 if the business relationship exists, but it is not so intense, and 1 if the business relationship exists and it is an intense one. A Confirmatory Factor Analysis (CFA) test showed that the aforementioned questions were highly correlated; therefore, they were combined into an index for business collaborations ranging from 0 to 1. This measure reflects both the intensity and diversity of the business collaborations.

### 3.2.4 Control variables

In order to achieve better results, those variables which may be influential in the analyses were controlled. At the individual level, there was control for age since it might affect how individuals expand their connections and obtain larger capitals which can be effective in their ventures (Semrau and Werner, 2014). Age, as a continuous variable was measured by the number of the years. Then, the second control variable was gender. The entrepreneurial literature has especially emphasised the difference between men and women in networking (Renzulli et al., 2000; Ashourizadeh and Schøtt, 2013) and business performance (Langowitz and Minniti, 2007). Gender was measured as a binary variable (1 for female and 0 for male). Furthermore, education was considered as a significant variable which affects both networking and entrepreneurial activities (Samuelsson and Davidsson, 2009; Semrau and Werner, 2014), and it was measured as a continuous variable via the number of years of education. Additionally, the perception of individuals about their skills and experience, fear of failure and identifying new opportunities in the market was also controlled (Peroni et al., 2016). Such variables were designed to be binary and were coded 1 for a 'Yes' and 0 for a 'No'. Finally, we also control the inclusion of entrepreneurial motives which show whether entrepreneurs were into entrepreneurship by necessity or by opportunity (Semrau and Werner, 2014). For modelling motivation including opportunity, necessity, opportunity and necessity, and also better opportunity, this categorical variable was re-coded into four dummies, taking the first category (i.e. opportunity) as the reference in order to compare other dummies to it.

At the firm level, previous studies revealed that firm characteristics such as size (Bonaccorsi, 1992), age and phase (Majumdar, 1997; Ling et al., 2007) are significantly effective on the performance of the firm. The firm age is a continuous variable achieved through subtraction of the current year and the year that the company was launched. Firm size, another continuous variable, was obtained through stating the number of staffs in the firm, excluding the owner. For both variables, age and size, we took a natural logarithm to reduce the skewness. Additionally, we controlled for the number of owners and sole ownership (Harrison and Freeman, 1999), since owners may bring financial and social capital to the firm which facilitate its performance. We also controlled the industry sector since some industries might be costly and have performance barriers.

Finally, since the economic conditions of each country are likely to influence a firm's exports (Hessels and Terjesen, 2010), the sample's data retrieved from the World Bank for Gross National Income (GNI) for this were controlled.

### 3.3 Method of analysis

Our data include entrepreneurs who were living in 71 different countries. Since the similarity of behaviour within each country is ignored and national-level factors are underestimated in multiple regression, we applied a hierarchical linear modelling test (Raudenbush and Bryk, 2002; Jensen and Schøtt, 2015) which considers both the similarity of behaviours within each society and number of the countries for the national factors. Although both multiple regression and the hierarchical linear modelling test estimate coefficients, the latter test includes fixed effects of independent variables and random effects of the countries, where it randomly selects a sample of countries (Raudenbush and Bryk, 2002). Six models were estimated for analysing the causal model in this study. In Model 1, the effect of the control variables was estimated. In Models 2 and 3, the independent variables were included, separately. In Model 4, we tested the effect of migration status on business collaborations as a dependent variable. In Model 5, we applied both independent variables simultaneously while the dependent variable was export. In Model 6, we tested the interaction effect of independent variables on the dependent variable.

## 4 Results

Analyses via the ANOVA test showed that the export average for immigrants (Mean = 18.04) was considerably higher than for natives (Mean = 8.28) and it was significant ($p < .000$) which was in line with a previous study on immigrants' export (Light and Shahlapour, 2016). Moreover, business collaborations were relatively higher among immigrants (Mean = .32) than natives (Mean = .24), using the ANOVA T-test ($p < .000$). Although the education rate was higher among immigrants than natives, the entrepreneurial competencies were similar for both groups. Comparing the firm's characteristics showed that the immigrants' firms were younger while the size was larger, and it was mostly established through joint ownership. Table 1 presents the descriptive analyses of the sample.

**Table 1**  Descriptive analyses of the sample

|  | *Immigrant* | | *Native* | |
| --- | --- | --- | --- | --- |
|  | *Mean* | *SD* | *Mean* | *SD* |
| Export | 18.04 | 29.03 | 8.28 | 20.28 |
| Business collaboration | .32 | .3109 | .24 | .3 |
| Age | 40.16 | 11.68 | 38.80 | 11.73 |
| Education | 12.55 | 3.91 | 10.61 | 4.67 |
| Opportunity seeking | .56 | .49 | .60 | .48 |
| Skills and experience | .84 | .36 | .83 | .37 |
| No fear of failure | .72 | .45 | .72 | .45 |
| Firm age | 1.06 | 1.17 | 1.17 | 1.17 |
| Firm size | .72 | 1.01 | .62 | .88 |
| Firm phase | .63 | .48 | .67 | .47 |
| Sole owner | .60 | .48 | .68 | .46 |
| N of owners | .94 | .39 | .88 | .35 |
| N | 2156 | | 33,080 | |

Table 2 presents the correlations among the variables. As expected, export had a positive correlation with business relations and migration status. Since business relations and migration status had a positive correlation, we tested their multicollinearity by the Variance Inflation Factor (VIF) (Stine, 1995). The correlations among independent variables were less than 2, suggesting that the multicollinearity was not an issue in this study (ibid).

Table 3 presents the results from the hierarchical linear model test. Model 1 included control variables, Models 2 and 3 examined the direct effect of business relations and migration status separately on export. In Model 4, the impact of migration status on business relations was tested while Model 5 was concerned with the concurrent direct effect of both independent variables on export. Finally, Model 6 considered the interaction effect of migration status and business relations upon export.

The first hypothesis H1 proposed the presence of more business relationships among immigrants than natives. Model 4 showed that being an immigrant had a positive significant impact on business relationships ($\beta = .01$, $p < .05$). In other words, immigrant entrepreneurs enjoyed a larger number of contacts with various business actors than native entrepreneurs. Thus, hypothesis H1 was supported.

**Table 2**  Mean, standard deviation and correlations among variables

| | | Mean | SD | 1 | 2 | 3 | 4 | 5 | 6 | 7 | 8 | 9 | 10 | 11 | 12 | 13 | 14 | 15 | 16 | 17 |
|---|---|---|---|---|---|---|---|---|---|---|---|---|---|---|---|---|---|---|---|---|
| 1 | Export | 8.92 | 21.11 | 1 | | | | | | | | | | | | | | | | |
| 2 | Business collab. | 0.25 | 0.30 | .18** | 1 | | | | | | | | | | | | | | | |
| 3 | Migration status | 0.07 | 0.25 | .11** | .07** | 1 | | | | | | | | | | | | | | |
| 4 | Age | 38.88 | 11.74 | –.02** | –.02** | .03** | 1 | | | | | | | | | | | | | |
| 5 | Education | 10.74 | 4.65 | .13** | .23** | .10** | –.06** | 1 | | | | | | | | | | | | |
| 6 | Gender (female = 1) | 0.42 | 0.49 | –.06** | –.10** | –.02** | –.01** | –.06** | 1 | | | | | | | | | | | |
| 7 | Opportunity Seeking | 0.60 | 0.49 | .01** | .01** | –.02** | –.14** | –.06** | .02** | 1 | | | | | | | | | | |
| 8 | Skills and experience | 0.83 | 0.38 | .04** | .05** | .01** | –.03** | .05** | –.04** | .16** | 1 | | | | | | | | | |
| 9 | Risk-willingness | 0.72 | 0.45 | .01** | .01** | .00** | –.04** | –.01** | –.03** | .14** | .18** | 1 | | | | | | | | |
| 10 | Motives | 6.18 | 3.68 | .04** | .09** | .01** | –.06** | .14** | –.04** | .09** | .06** | .08** | 1 | | | | | | | |
| 11 | Firm age | 1.17 | 1.18 | –.08** | –.04** | –.02** | .36** | –.08** | –.04** | –.18** | –.07** | –.06** | –.06** | 1 | | | | | | |
| 12 | Firm size | 0.63 | 0.90 | .11** | .21** | .03** | .09** | .14** | –.11** | –.04** | .02** | .00** | .04** | .33** | 1 | | | | | |
| 13 | Firm phase | 0.67 | 0.47 | –.07** | –.01** | –.02** | .20** | –.06** | –.03** | –.11** | –.04** | –.04** | –.04** | .71** | .51** | 1 | | | | |
| 14 | Sole owner | 0.68 | 0.47 | –.01** | –.15** | –.04** | .09** | –.16** | .04** | .02** | .03** | .02** | –.05** | .12** | –.11** | .12** | 1 | | | |
| 15 | Number of owners | 0.89 | 0.38 | .11** | .16** | .04** | –.06** | .15** | –.06** | –.02** | –.03** | –.01** | .05** | –.08** | .18** | –.08** | –.80** | 1 | | |
| 16 | Industry sector | 3.20 | 1.03 | –.03** | –.06** | 0.0** | –.09** | .01** | .20** | .08** | .05** | .02** | .03** | –.12** | –.07** | –.07** | .04** | –.05** | 1 | |
| 17 | GNI | 12.78 | 10.91 | .10** | .15** | .22** | .24** | .30** | –.07** | –.22** | –.04** | –.04** | .04** | .15** | .08** | .05** | –.11** | .10** | –.10** | 1 |

Notes: *Correlation is significant at .05 (two tailed).
**Correlation is significant at .01 (two tailed).

*Immigrant and native's export benefiting from business collaborations*

The second hypothesis H2 was about the positive effect of business collaborations on export. Models 2 and 5 confirmed that the business collaborations promoted exporting among all the entrepreneurs (Model 2: $\beta = 6.85$, $p < .000$ and Model 5: $\beta = 6.82$, $p < .000$). It meant that attaining a greater number and more business collaborations did promote export. Therefore, hypothesis H2 was supported by these tests.

Models 3 and 5 confirmed previous studies (Ashourizadeh et al., 2016; Light and Shahlapour, 2016) which found that migration status had a significant positive effect on export performance (Model 3: $\beta = 3.85$, $p < .000$ and Model 5: $\beta = 3.77$, $p < .000$).

Hypothesis H3 assumed an enhancing moderating effect of migration status upon collaborative business networking on export performance. Testing this hypothesis in Model 6, the coefficient was positive and very significant ($\beta = 6.3$, $p < .000$). It meant that being an immigrant promoted the positive influence of business relations on export performance.

**Table 3** The effect of business collaborations on export via hierarchical linear model test

|  | Model 1 | Model 2 | Model 3 | Model 4# | Model 5 | Model 6 |
|---|---|---|---|---|---|---|
| Business collaborations |  | 6.85*** |  |  | 6.82*** | 6.3*** |
| Migration status (immigrant = 1) |  |  | 3.85*** | .01* | 3.77*** | 1.6** |
| Business relations*migration status |  |  |  |  |  | 6.7*** |
| Age | .0008 | .001 | –.0007 | –9.64 | 2.46 | –.0002 |
| Education | .14*** | .106*** | .14*** | .006*** | .09*** | .1*** |
| Gender (female = 1) | –.96*** | –.782*** | –.94*** | –.02*** | –.76*** | –.75** |
| Opportunity seeking | .19 | .0009 | .19 | .02*** | –.004 | –.007 |
| Skills and experience | .61* | .39 | .61* | .03*** | .401 | .4 |
| No fear of failure | .03 | .03 | .04 | –.0008 | .03 | .04 |
| Motive (necessity) | –1.36*** | –1.03*** | –1.37*** | –.04*** | –1.05*** | –1.05*** |
| Motive (opportunity and necessity) | –.92** | –.807* | –.95** | –.01*** | –.83* | –.83* |
| Motive (better opportunity) | –.95* | –.94* | –.92* | .0005 | –.91* | –.91* |
| Motive (other) | –.66 | –.52 | –.69 | –.01** | –.55 | –.52 |
| Firm age | –.58*** | –.47*** | –.55*** | –.01*** | –.45** | –.44** |
| Firm size | 2.24*** | 1.81*** | 2.23*** | .06*** | 1.79*** | 1.8*** |
| Firm phase | –2.52*** | –2.34*** | –2.53*** | –.02*** | –2.36*** | –2.3*** |
| Sole owner | .66* | .718196* | .68* | –.01* | .74* | .75* |
| N of owner | 3.29*** | 2.95*** | 3.31*** | .05*** | 2.97*** | 2.9*** |
| Industry sector | –.25* | –.21* | –.26* | –.006*** | –.22* | –.22* |
| GNI | .19** | .17** | .16** | .002** | .15** | .15** |
| Constant | 4.66** | 3.72* | 3.85*** | .13*** | 3.84* | 3.94** |
| Random effect | Yes | Yes | Yes | Yes | Yes | Yes |

Notes: *, ** and *** denote level of significance at .05, .005 and .0005.

*4.1 Robustness test*

In order to make sure that the obtained results were reliable, we ran a robustness test in this study. Splitting immigrants from natives in Models 7 and 8, the results revealed that immigrant's stronger business collaborations (Model 7: $\beta$ = 12.36, $p$ < .000) than natives (Model 8: $\beta$ = 6.45, $p$ < .000) and it positively affected their export performance. Therefore, our hypotheses were confirmed through robustness checks. Table 4 presents the results.

**Table 4** Additional robustness checks

|  | Model 7 | Model 8 |
|---|---|---|
|  | *Immigrant* | *Native* |
| Business collaboration | 12.36*** | 6.45*** |
| Age | −.03 | .003 |
| Education | .16 | .1*** |
| Gender (female = 0) | −.89 | −.8*** |
| Opportunity seeking | 1.42 | −.19 |
| Skills and experience | .29 | .41 |
| No fear of failure | −.23 | .07 |
| Motive (necessity) | −2.34 | −.1*** |
| Motive (opportunity and necessity) | 1.14 | −.94** |
| Motive (better opportunity) | 2.09 | −1.07* |
| Motive (other) | −1.35 | −.37 |
| Firm age | .24 | −.49*** |
| Firm size | 2.39** | 1.8*** |
| Firm phase | −6.31** | −2.13*** |
| Sole owner | 1.46 | .64* |
| N of owner | 7.09** | 2.62*** |
| Industry sector | −.12 | −.22* |
| GNI | .11 | .15** |
| Constant | 4.72 | 4.07** |
| Random effect | Yes | Yes |

Notes: *, ** and *** denote level of significance at .05, .005 and .0005.

**5 Discussion and conclusion**

In the beginning of this paper, the authors raised two questions regarding the impact of migration on entrepreneurs' business collaborations and its benefits for export: *How different* is the business collaboration network among immigrant entrepreneurs as compared to the natives? How different is the benefit of business collaboration network for exporting among immigrant entrepreneurs as compared to the natives? In the following, the results of hypothesis test for the two questions has been provided. Further discussion on the results has been presented.

## 5.1 Discussion of findings

Applying the theories of social network (Granovetter, 1985; Burt, 2001) and immigrant entrepreneurship (Light et al., 1993; Robertson and Grant, 2016), we proposed that immigrant entrepreneurs had more business collaborations than native entrepreneurs. Controlling influencing factors at individual, firm and national level, we found that immigrant entrepreneurs held more business collaborations compared to native entrepreneurs. This finding concurs with prior comparative studies on immigrant' networks (e.g. Light and Shahlapour, 2016; Moghaddam et al., 2017) and may arise from the nature of the immigrants' dual embeddedness (Drori et al., 2009) that encourages them to be in a brokerage situation between two unrelated networks of home and host country. Consequently, they attain more business collaborations than native entrepreneurs.

Additionally, we examined the impact of such business collaborations on immigrant entrepreneurs' export when compared to that of native entrepreneurs. The results turned out to be very significant for the immigrant entrepreneurs in the way they benefited more from their business collaborations for exporting than did the native entrepreneurs. Thus, the findings coincide with those of previous research which consider the effect of business collaborations on innovation as embedded in the firm's network of collaborations (Hite, 2005; Jensen and Schøtt, 2014; Schøtt and Sedaghat, 2014) in the sense that the wider business collaborations can facilitate innovations (Jensen and Schøtt, 2015) and more innovations can lead to the growth and internationalisation of the firms (Roper and Love, 2002; Saeedikiya et al., 2017).

In addition to this, immigrant entrepreneurs tend to have an informational advantage over the natives in that they receive information and advice from heterogeneous sources (Drori et al., 2009; Griffin-El and Olabisi, 2018). The indication is that different ideas for export can be received from various perspectives. Another explanation could be the situational advantage of the immigrant entrepreneurs over the natives in that they are likely to have a comparative edge over the natives by knowing the specific needs and possibilities of the two societies (Kloosterman, 2010; Aleksynska and Peri, 2014). Such knowledge allows them to benefit from their networks for export. Our findings support previous studies on the potential of immigrants to perform better than native entrepreneurs in their business activities (Drori et al., 2009; Ashourizadeh et al., 2016; Efendic et al., 2016). The results also support the idea that an immigrant entrepreneur can more easily identify and exploit cross-border opportunities from their home and host countries (Emontspool and Servais, 2017; Griffin-El and Olabisi, 2018).

## 5.2 Theoretical contribution

The contribution of this study to the subject is three-fold. First, the current study helps to address the paucity of research on immigrants' business collaborations and their impact on their entrepreneurial activities. While previous studies examined the benefits of ethnic and kin networks to immigrants (e.g. Schrover et al., 2007) or native entrepreneurs (Granovetter, 1985) for their business, this study investigated the advantages of business collaborations for immigrant entrepreneurs and their export activities in their host countries when compared to those of their native counterparts. Secondly, the findings also contribute to the existing theory on how the dual embeddedness of immigrants' networking in their home and host countries affect their social capital benefit (Rath and Kloosterman, 2000; Rusinovic, 2008; Drori et al., 2009; Patel and Conklin, 2009) for exporting. Through dual embeddedness and their strategic position which allow them to operate in the two different contexts, immigrant entrepreneurs have access to more social capital, such as knowledge and information regarding the international market (Drori et al., 2009; Griffin-El and Olabisi, 2018). Hence, they tend to have better opportunities to combine information from different sources to develop innovative ideas for expanding their business, especially in the international market.

Thirdly, compared to previous studies that investigated immigrant entrepreneurs either through the census data of a country, such as the USA (Mora and Dávila, 2006; Wadhwa et al., 2007), or from self-collected data (Raijman and Tienda, 2000; Collins and Low, 2010) limited to one or few specific ethnic communities (Portes, 1987; Light and Bonacich, 1991; Andersson and Hammarstedt, 2011), our study presents a more comprehensive picture of the reality due to the large number of countries included. The scope is wide because the researchers used a cross-sectional approach to compare data from 47,200 immigrant and native entrepreneurs in 71 countries. This was done in order to investigate the benefits of business collaborations for export as a way of internationalising their enterprises. While we treat the context of the countries as an analytical variable (Cheng, 2007) which we controlled, the cases in this data were chosen randomly. Therefore, the findings can be generalised and applied to the population of immigrant and native entrepreneurs worldwide.

### 5.3 Managerial and policy implications

A 51% increase in international migration during the last 15 years has raised concerns and generated many discussions in the media in the USA and in Europe (*Daily Telegraph*, 2016; *The Wall Street Journal*, 2016). Therefore, the findings of this study have timely implications for managers and policy-makers in Europe and the USA. The findings imply that immigrants possess great potential for boosting national economies (Peri, 2012; United Nations, 2016) and business collaborations is one of the effective ways to improve their international activities, like export. Accordingly, it is being recommended that European governments devise integrating programs which provide a platform for immigrant and native entrepreneurs to meet and exchange knowledge and information with each other. Immigrant entrepreneurs with diverse and strong networks in different societies can become potential business collaborators for native entrepreneurs. Providing subsidies for immigrant entrepreneurs to participate in international exhibitions which native entrepreneurs attend or connecting immigrant entrepreneurs with members of chambers of commerce in the host country is another way to improve business collaboration between immigrant and native entrepreneurs and, consequently, international performance between both types of entrepreneurs.

### 5.4 Limitations and future studies

Although the comprehensiveness of the data enables the generalisation of the results, it has limited the depth of study. For instance, while the immigrants' business collaborations have been measured, the location of the networks is unknown. Knowing the location of these business collaborations would provide a better explanation of why immigrant entrepreneurs export to a specific destination and why they perform as they do in a specific type of industry. Additionally, it is being recommended that there should be a closer examination of the economic, regulatory and cultural institutions of the target countries and how these various factors might affect the business collaborations and export performance among immigrant and native entrepreneurs. An example of one such study is that done by Wales et al. (2016) which showed that a strong relationship with government in an emerging economy is more significant than the regulative context of the society. Hence, it presents an interesting comparison between developing and developed countries from an institutional viewpoint.

It has been argued that network density plays an important role in the firm's success and performance (Birley et al., 1991; Meagher and Rogers, 2004). Therefore, the undertaking of future research is being proposed to address this issue in terms of providing a comparative insight of immigrant and native entrepreneurs' network density and its effect on export. Finally, a study of the characteristics of the immigrants' firms, which were considered in this study, is necessary based on the finding that the joint venture type of firm is more prevalent among immigrant entrepreneurs and the number of owners was also higher. The larger number of owners from the social capital perspective might be more effective on the firm's performance, especially in the international arena. Therefore, it should be of some concern as to how a firm's characteristics might influence the nature of its interactions within these business collaborations as well as the export performances of immigrant compared to native entrepreneurs.

## Acknowledgements

We thank the anonymous reviewers for their careful reading of our manuscript and their many insightful comments and suggestions. This paper is supported by the National Natural Science Foundation of China (71772103) and Tsinghua University Institute for China Sustainable Urbanization (TUCSU-K-17024-01). Data are collected by Global Entrepreneurship Monitor and responsibility of the statistical analyses remains to the authors.